# Different strategies for cancer treatment: mathematical modeling

O.G. Isaeva [a,*] and V.A. Osipov [a]

[a] *Bogoliubov Laboratory of Theoretical Physics, Joint Institute for Nuclear Research, 141980, Dubna, Moscow region, Russia*

**Abstract**

We formulate and analyze a mathematical model describing immune response to avascular tumor under the influence of immunotherapy and chemotherapy and their combinations as well as vaccine treatments. The effect of vaccine therapy is considered as a parametric perturbation of the model. In the case of a weak immune response, neither immunotherapy nor chemotherapy is found to cause tumor regression to a small size, which would be below the clinically detectable threshold. Numerical simulations show that the efficiency of vaccine therapy depends on both the tumor size and the condition of immune system as well as on the response of the organism to vaccination. In particular, we found that vaccine therapy becomes more effective when used without time delay from a prescribed date of vaccination after surgery and is ineffective without preliminary treatment. For a strong immune response, our model predicts the tumor remission under vaccine therapy. Our study of successive chemo/immuno, immuno/chemo and concurrent chemoimmunotherapy shows that the chemo/immuno sequence is more effective while concurrent chemoimmunotherapy is more sparing.

---

[*] Corresponding author. Current addresses: Bogoliubov Laboratory of Theoretical Physics, Joint Institute for Nuclear Research, 141980, Dubna, Moscow region, Russia. Fax.: (7)(49621)65084. E-mail addresses: issaeva@theor.jinr.ru, osipov@theor.jinr.ru

# 1. Introduction

Some modern trends in treatment of cancer are based on the ability of certain forms of tumors to stimulate immune response. The fact that the immune system plays an important role in fighting cancer has been verified both in laboratorial and clinical experiments [15,34]. An inclusion of immune component in mathematical models of tumor growth has been shown to reflect clinically observed phenomena such as uncontrolled growth of tumor, tumor dormancy and oscillations in tumor size [1,8,10,11,16,24,25,44,47]. A similar tumor behavior was also predicted in our recent ODE model [22] with the interleukine-2 (IL-2) taken into account.

The goal of immunotherapy is to enhance the anti-tumor resistance of an organism and improve the immune system condition. There are known three main categories of immunotherapy: immune response modifiers (cytokines), monoclonal antibodies and vaccines [40]. Such immune modifiers as IL-2, interferon-$\alpha$ (IFN–$\alpha$) as well as tumor necrosis factor-$\alpha$ (TNF–$\alpha$) are already widely used in cancer immunotherapy [4,5,19,20,41,48]. An important problem is to choose the correct schedule for using chemotherapy in combination with IL-2 and IFN–$\alpha$ therapy. For example, a series of sequential phase II trials based on integrating of IL-2 and IFN–$\alpha$ with the CVD (cisplatin, vinblastine, and dacarbazine) regimen shows that chemotherapy followed immediately by immunotherapy is more effective for treatment of human metastatic melanoma than their reverse sequence [4,5]. It was also observed that concurrent chemoimmunotherapy is almost as effective as chemo/immune sequence when immunotherapy is administered right after the CVD. At the same time, the concurrent chemoimmunotherapy is found to be less toxic than the sequential regimens [4]. Monoclonal antibodies (MA) are used in both diagnostics and therapy of cancer. This follows from ability of MA to recognize tumor antigens on a surface of tumor cells. As a result, MA can deliver both anti-tumor drugs and radioactive isotopes exactly to the malignant cells [27,37]. In spite of the fact that cancer vaccines are still under experimental investigations, the existent clinical trials clearly show that they can improve immune response to certain forms of cancer [40,50]. Most of cancer vaccines consist of living tumor cells and their lysis products while some of them contain tumor-derived proteins, peptides and gangliosides [42]. For instance, an experimental vaccine for malignant melanoma consists of four melanoma peptides restricted by HLA-A1, A2, A3 and HLA-DR and includes IL-2 and granulocyte macrophage colony stimulating factor (GM-CSF) as adjuvants. It was found that this vaccination is able to stimulate tumor regression in some cases [45,46]. The observed toxicity of this vaccination is connected with low doses of IL-2. These findings stimulated our interest to consider within our model the effects of combination immune and chemotherapy treatments as well as vaccine therapy.

One of the first attempts to consider effects of immunotherapy within an appropriate ODE model was made by Kirschner and Panetta in [24]. They study immunotherapy based on the use of IL-2 together with adoptive cellular immunotherapy (ACI) by introducing in dynamical equations terms describing external inflow of both IL-2 and cultured immune cells. More recently, de Pillis and Radunskaya have proposed the kinetic model of anti-tumor immune response where individual equations were suggested for the description of mechanisms of natural immunological defense presented by NK-cells and specific immune response presented by CD8+ T cells [10,11]. Notice that unlike [24] they do not consider a natural dynamics of IL-2. In the framework of this model the effects of chemotherapy, immunotherapy, their combined influence, as well as the vaccine therapy were considered [11].

It is interesting to mention a recent paper by Arciero et al. [2] where a novel treatment strategy known as small interfering RNA (siRNA) therapy was considered in the framework of the model proposed in [24]. This treatment suppresses TGF-β production by targeting the mRNA codes for TGF-β, thereby reducing the presence and effect of TGF-β in tumor cells. The model predicts conditions under which siRNA treatment can be successful in transformation of TGF-β producing tumors to either non-producing or producing a small value of TGF-β tumors, that is to a non-immune evading state.

In recent years, the importance of spatial aspects of tumor-immune dynamics was demonstrated [30-32,35,36]. For example, in [31,32] the tumor cell distributions that are quasi-stationary in time and heterogeneous in space were studied within the PDE model by Matzavinos and Chaplain based on the ODE model by Kuznetsov [25]. It was found that depending on model parameters this reaction-diffusion-chemotaxis system is able to simulate the well-documented phenomenon of cancer dormancy, as well as tumor invasion which is presented in the form of a standard traveling wave. The interesting PDE model of non-specific immune response was developed by Owen and Sherrat in [35,36]. The important conclusion of this model is that macrophages are unable to prevent tumor growth. Nevertheless, significant effects on the form of the tumor were predicted, including the formation of spatial patterns. The model demonstrates the existence of traveling wave solutions connecting the normal tissue and tumor steady states corresponding to a growing tumor. When macrophage chemotaxis is included, these patterns can in some cases bifurcate to give irregular spatiotemporal oscillations.

Finally, let us mention the hybrid cellular automata–PDE modeling approach which combines continuous PDEs for chemical quantities and a discrete cell-based description for biological cell species with phenomenologically sourced probabilities for cell dynamics [30]. This approach allows one to consider both temporal and two-dimensional spatial evolution of the system. Numerical simulations include spherical tumor growth, stable and unstable oscillatory tumor growth, satellitosis and tumor infiltration by immune cells.

Our model of immune response to early (avascular) tumor growth is based on the mechanism of intercellular cytokine mediated interaction in cellular immune response proposed by Wagner et al. [49] which was modified by taking into account co-stimulatory factors (see, e.g., [28,39]). Generally, it consists of seven ordinary differential equations. To simplify analysis, we reduced the model to three equations incorporating the most important modern concepts of tumor-immune dynamics including the influence of IL-2 dynamics (see [22] for details). Notice also that we do not consider the spatial migration of cell populations that is of most importance for modeling of angiogenesis (vascular growth), invasion and metastasis (see, e.g., [7]).

In this paper, we extend our model [22] to describe chemo- and immunotherapy effects. The outline of the paper is as follows. First of all, a mathematical model of tumor-immune dynamics under the influence of both immunotherapy with IL-2 and IFN−α and chemotherapy is formulated in section 2. In section 3 we perform a steady state analysis of the model. The results of numerical studies are presented in section 4 for four different cases: chemotherapy alone, IL-2 alone, IL-2 plus IFN−α, and a combination of chemotherapy and immunotherapy (IL-2 therapy). The effects of vaccine therapy are considered in the absence of chemotherapy and immunotherapy. Section 5 is devoted to conclusions and discussion.

## 2. Mathematical model

The system (1)—(5) describes the most important components of tumor-immune dynamics in the presence of treatment components. Namely, we consider five populations: tumor cells (*T*), CTL (*L*), IL-2 ($I_2$), chemotherapeutic drug (*C*), and IFN−α (*I*).

$$\frac{dT}{dt} = -aT \ln \frac{bT}{a} - c(I)TL - M_T(I_2)(1 - e^{-C})T, \quad (1)$$

$$\frac{dL}{dt} = d + eLI_2 - fL - M_L(I_2)(1 - e^{-C})L, \quad (2)$$

$$\frac{dI_2}{dt} = V_{I_2}(t) + \frac{gT}{T+l} - jLI_2 - kTI_2, \quad (3)$$

$$\frac{dC}{dt} = V_C(t) - pC, \quad (4)$$

$$\frac{dI}{dt} = V_I(t) - qI. \quad (5)$$

The tumor growth is described by the Gompertzian law (the first term in (1)). The destruction of tumor cells by CTL is presented by the second term in (1). It is supposed that the destruction rate is proportional to the number of tumor cells and CTL populations. In (2) *d* characterizes the steady inflow of CTL into the tumor site. Second and third terms in (2) describe CTL proliferation in response to the IL-2 action and CTL death rate, correspondingly. In (3)—(5) $V_i$ (*i*=$I_2$, *C*, *I*) describes the external influxes of IL-2, chemotherapeutic drug and IFN−α, respectively. Since therapy is assigned to a certain schedule, these influxes are taken to be time-dependent. IL-2 production in (3) is described by hyperbola (the second term), which allows us to take into account a limitation in the stimulation of the immune system by the growing tumor. At small *T* the growth rate is nearly linear in tumor size while for big tumor (*T* >> *l*) it tends to a maximum constant value *g*. The parameter *l* influences the IL-2 production rate. The smaller is the value of *l*, the quicker the IL-2 production rate achieves its maximum value *g*. Notice that *g* characterizes the degree of expression of the antigen plus major histocompatibility complexes class II (AG-MHC-II) on the APC surfaces, i.e. the antigen presentation. The probability of activation (provoking the IL-2 production) of helper T cell precursor by the APC increases with the antigen presentation. Since IL-2 is a short-distance cytokine, it is suggested that target cells (cytotoxic T lymphocytes) effectively consume IL-2. The consumption rate is presented by the third term in (3). It was found that inhibition of IL-2 results from an accumulation of immune-suppressing substances, prostaglandins. Their number is proportional to the concentration of tumor cells. Prostaglandins suppress the production of IL-2 and can directly destroy its molecules [38]. In (3) the IL-2 destruction rate is described by the fourth term.

The interaction of chemotherapeutic drug with sensitive cells can be described by using of either the Michaelis-Menten kinetics [23] or the exponentially saturating kinetics [11]. Similarly to [11], we use in Eqs. (1) and (2) a saturation term $M_j(I_2)(1-e^{-C})j$ with $j = T, L$ to describe cell death caused by chemotherapeutic drug. At low concentrations the death rate is nearly linear in drug while at higher concentrations the death rate turns out to be $C$-independent. As was noted in [11] this behavior shows a good correlation with existing dose-response curves (see, e.g., [18]). We assume that $M_j$ depends on the IL-2 concentration in the following way: $M_j(I_2) = M_j^{\text{chemo}}(2 - e^{-I_2/I_{20}})$. Thus, $M_j$ increases with the concentration of IL-2, however, it never exceeds a doubled value of $M_j^{\text{chemo}}$ (cell killing by chemotherapy). This is based on the fact that IL-2 can induce the secondary cytokines such as TNF-$\alpha$, which could enhance the anti-tumor effect of cytotoxic chemical agents (see, e.g., [5]). We also suppose that the model parameter $c$ in (1) depends on the IFN−$\alpha$ concentration as $c(I) = c_{\text{CTL}}(2 - e^{-I/I_0})$, where $c_{\text{CTL}}$ is a rate of tumor cells inactivation by CTL. This agrees with the fact that IFN−$\alpha$ enhances immune-mediated anti-tumor responses by increasing expression of MHC molecules on tumor cells, thus enhancing their recognition by CTL [48]. Notice that only therapeutic IFN−$\alpha$ dynamics is considered within our model.

## 2.1. Parameter set

As is shown in our previous work [22], possible scenarios of tumor-immune dynamics are very sensitive to the choice of the parameters in equations (1)—(5). In fact the parameter sets vary not only for specific cancer types but also from one individual to another. Our model is based on using of some generalized (most typical) parameters. In order to reflect the individual clinical outcomes we conditionally divide patients in three groups (see Tables 1 and 2). We assume that tumor has the same histological structure (for instance, melanoma) with equal doubling time and carrying capacity (actually, these characteristics may vary between tumor specimens). Additionally, the lifetime of CTL is chosen to be the same. On the other hand, the tumor antigen expression ($c_{CTL}$), the strength of the immune response ($e$, $g$, $j$ and $k$), and the reaction to vaccination are taken to be specific for each group. Some values of model parameters were estimated by using the available experimental data. In particular, the human melanoma growth parameters $a$ and $b$ were obtained from the experimental data found in Hu's results on mice trials where human melanoma was tested in a severe combined immunodeficient mice [21]. Using the least-squares method, we fitted the experimental curve produced by the data of a control group to Gompertzian curve. The death rate of CTL was estimated using the relation $f = 1/\tau$ where $\tau$ is their known average lifetime. The rate of steady inflow of CTL was calculated from the relation $d = fL_{free}$ where $L_{free}$ (the number of CTL capable to recognize melanoma specific antigen in the organism without tumor) was estimated to be about $2.25\times10^7$ cells using the data for the full number of CD8+ T cells in blood and a percent value of T cells specific for melanoma antigen [17]. The parameters characterizing the cell death caused by chemotherapeutic influence $M_T^{chemo}$ and $M_L^{chemo}$ are taken from de Pillis's model [11]. The elimination rates for chemotherapeutic drug (dacarbazine) and IFN−α were estimated by using their known half-life times and the relations: $p = \ln2/t_{C1/2}$ and $q = \ln2/t_{I1/2}$ (see, for example, [29,43]). For the rest of parameters we chose values most appropriate to our model. Current medical literature and sensitivity analysis (see [22]) allow us to conclude that the corresponding interactions are of importance in the description of immune response.

# 3 Non-dimensionalization, steady state analysis

## 3.1. Scaling

For convenience let us introduce dimensionless variables and parameters as follows: $T' = T/T_0$, $L' = L/L_0$, $I'_2 = I_2/I_{20}$, $I' = I/I_0$ and $t' = t/\tau$, where $\tau = f^{-1}$ (days). The values of $T_0$, $L_0$, $I_{20}$ and $I_0$ are given in accordance with [24] and presented in Table 1. Notice that the variable for chemotherapeutic drug, $C$, is given in relative units. The choice of the time-scale factor $\tau$ is based on the fact that the mean lifetime of CTL is about three days and a similar time is needed for the proliferation of CTL and IL-2 production [6,9].

Dropping primes for notational clarity, equations (1)—(5) take the following form in normalized units:

$$\frac{dT}{dt} = -h_1 T \ln \frac{h_2 T}{h_1} - h_3(2 - e^{-I})TL - m_1(2 - e^{-I_2})(1 - e^{-C})T, \quad (6)$$

$$\frac{dL}{dt} = h_4 + h_5 LI_2 - L - m_2(2 - e^{-I_2})(1 - e^{-C})L, \quad (7)$$

$$\frac{dI_2}{dt} = m_3(t) + \frac{h_6 T}{T + h_9} - h_7 LI_2 - h_8 TI_2, \quad (8)$$

$$\frac{dC}{dt} = m_4(t) - m_5 C, \quad (9)$$

$$\frac{dI}{dt} = m_6(t) - m_7 I, \quad (10)$$

where $h_1 = a/f$, $h_2 = bT_0/f$, $h_3 = c_{CTL}L_0/f$, $m_1 = M_T^{chemo}/f$, $h_4 = d/fL_0$, $h_5 = eI_{20}/f$, $m_2 = M_L^{chemo}/f$, $m_3(t) = V_{I_2}(t)/fI_{20}$, $h_6 = g/fI_{20}$, $h_7 = jL_0/f$, $h_8 = kT_0/f$, $h_9 = l/T_0$, $m_4(t) = V_C(t)/f$, $m_5 = p/f$, $m_6(t) = V_I(t)/fI_0$, and $m_7 = q/f$.

## 3.2. Steady state analysis

To perform a steady state analysis we study the system

$$\frac{dT}{dt} = -h_1 T \ln \frac{h_2 T}{h_1} - h_3 TL, \quad (11)$$

$$\frac{dL}{dt} = h_4 + h_5 L I_2 - L, \tag{12}$$

$$\frac{dI_2}{dt} = \frac{h_6 T}{T + h_9} - h_7 L I_2 - h_8 T I_2, \tag{13}$$

which follows from (6)—(10) at $V_i(t) = 0$ ($i = I_2, C, I$) and $C(0) = I(0) = 0$.

A possible way to perform the steady state analysis is to use isoclines. Let us consider the phase plane $TL$, which shows the interactions between two main cell populations: tumor cells and CTL. In this case, the equations for horizontal and vertical isoclines are written as

$$(h_4 - L)(T + h_9)(h_7 L + h_8 T) + h_5 h_6 T L = 0, \tag{14}$$

$$L = -\frac{h_1}{h_3} \ln \frac{h_2 T}{h_1}, \quad T = 0. \tag{15}$$

The fixed points are situated at the intersections of isoclines (14) and (15). Our analysis shows that at any choice of parameters the system (11)—(13) has the unstable point (0, $h_4$, 0), which lies at the intersection of isoclines (14) and $T = 0$. This means that the regime of full tumor regression is not allowed.

We consider $g$ (characterizing the antigen presentation) as a varying parameter. A bifurcation diagram for the dimensionless parameter $h_6$ is presented in Figure 1 where the function $h_6(T)$ is obtained by substitution of $L$ from (15) into (14). As is seen, there are two bifurcation points. Therefore one can distinguish three main dynamical regimes. The *region* I ($h_6 < h_{6min}$) characterizes the weak immune response. The system (11)—(13) has two fixed points: a saddle point (0, $h_4$, 0) and an improper node ($T_3$, $L_3$, $I_{23}$). This means that under a deficiency in the production of IL-2, the population of tumor cells is able to escape from the immune response. The tumor grows and the immune system becomes suppressed. In the *region* II ($h_{6min} < h_6 < h_{6max}$), which we associate with the strong immune response, there appear two additional fixed points: a stable spiral ($T_1$, $L_1$, $I_{21}$) and a saddle ($T_2$, $L_2$, $I_{22}$). Therefore different regimes can exist depending on the initial conditions. First, when initial CTL population size is sufficiently large to reduce a tumor population, the regression of tumor up to a small fixed size where the dynamical equilibrium between tumor and immune system is reached. In this case, the tumor manifests itself via the excited immune system. Second regime appears when initial number of CTL is not large enough to drive the system at the dynamical equilibrium point ($T_1$, $L_1$, $I_{21}$), which is a stable spiral. Thus, the tumor grows to a highest possible size, which is defined for the tumor population being in conditions of restricted feeding. The dynamical equilibrium between the tumor and the immune system is reached at the fixed point ($T_3$, $L_3$, $I_{23}$) that is an improper node. Finally, in the *region* III ($h_6 > h_{6max}$) the fixed points ($T_2$, $L_2$, $I_{22}$) and ($T_3$, $L_3$, $I_{23}$) disappear. As a result, there are two fixed points: a saddle point (0, $h_4$, 0) and a stable spiral ($T_1$, $L_1$, $I_{21}$). In this case, a decrease in tumor size is found when the equilibrium between the tumor and the immune system is established. The region III is associated with the dormant tumor when the immune system is able to handle the tumor size.

## 4 Numerical experiments

In this section we study the effects of chemotherapy alone, IL-2 alone, IL-2 plus IFN-α therapy, regimens of sequential chemoimmunotherapy as well as vaccine therapy for three groups of patients. Two of them (P1 and P2) generate weak immune responses to the tumor while the third one (P3) generates a strong immune response. The group P2 is characterized by a lower antigen expression in comparison with P1 and, correspondingly, exhibits a weaker immune response (see Tables 1 and 2). At the stage II of malignant melanoma both chemotherapy and immunotherapy are usually administered after surgical treatment. Therefore, the initial tumor size is assumed to take a hypothetical value of $T(0) \sim 8 \times 10^6$ cells. When we study the effects of treatment administered without preliminary surgery, the initial tumor size is assumed to take a hypothetical value of $3 \times 10^7$ cells. In subsections 4.1—4.6 we will consider the first group of patients (P1).

### *4.1. Chemotherapy*

Let us test a treatment approach which employs nine pulsed doses of chemotherapy, each dose represented by setting $V_C(t) = 1$ in (4) for a day, and given once every 5 days (Figure 2(d)). As is seen from Figure 2(a), a regression is not observed and the tumor population grows. The number of tumor cells oscillates in time as a result of pulsed character of dosing. Tumor growth rate is found to decrease in comparison with the case without treatment. This is completely due to chemotherapeutic influence because the CTL dynamics is slightly affected by chemotherapy (see Figure 2(b)). A possible reason is that an increase in CTL proliferation caused by increasing IL-2 concentration is compensated by death of CTL under the action of chemo-drug. Thus, our study shows that chemotherapy results in stunted tumor growth. In particular, at our choice of parameters the tumor achieves its dangerous size about ten days later than in the absence of the therapy.

### *4.2. Immunotherapy*

*IL-2 alone*

The following regimen of the IL-2 alone therapy is supposed: four pulsed doses of IL-2, each is equal to 10 MU/day for four days, and administered every 10 days. As is seen from Figure 2(a), there is a tumor remission with the duration of about 40 days. At the same time full tumor regression is not observed. Indeed, as IL-2 concentration grows, the CTL population is also increased approximately by a factor of 7 in 40 days (see Figure 2(b)). However, approximately ten days after treatment cessation the IL-2 concentration decreases (see Figure 2(c)). Accordingly, the CTL population also regresses and, as a result, the tumor growth revives. Thus, this course of treatment leads to a temporary remission only (for 1—1.5 months in our case).

*IL-2 plus IFN-$\alpha$*

Let us consider a combined course of the immunotherapy, when IL-2 and IFN-$\alpha$ are given simultaneously. The dose administration pattern for IL-2 is considered to be the same as in the previous subsection. Together with IL-2 the IFN-$\alpha$ at the dose 5 MU/day for four days in a 10 day cycle is administered (Figure 2(d)). As is shown in Figure 2(a), there is a substantial decrease in the number of the tumor cells during the cure. The tumor remission becomes more pronounced in comparison with the previous case although the regression time is almost the same.

Thus, our study shows that immunotherapy is more effective in the remission time of the tumor as compared with chemotherapy. As another conclusion, the IL-2 alone therapy should be considered as more sparing treatment in comparison with the case of IL-2+IFN-α. Indeed, in spite of better tumor remission for IL-2+IFN-α treatment the IL-2 alone therapy is less toxic.

*4.3. Sequential chemo/immunotherapy*

In the next three subsections, we study the effects of chemotherapy followed immediately by immunotherapy or vice versa, as well as the concurrent chemoimmunotherapy. We consider the following sequential therapy regimen: one pulse of chemotherapy is presented by setting in (4) $V_C(t) = 1$ per day for four days (Figure 3(d)). During the next four days one pulse of IL-2 therapy is administered in amounts of $V_{I_2}(t) = 10$ MU/day in (3). Figure 3(a) shows the dynamics of tumor cells. As is seen, the chosen regimen of sequential therapy does not lead to the tumor regression. However, a markedly stunted tumor growth is observed (tumor cell population reaches the maximum value about thirty days later in this case). At the initial stage ($t < 8$ days), the tumor growth deceleration is entirely due to chemotherapeutic impact. Furthermore, the tumor cell population slightly decreases. This effect is caused by an increase of the IL-2 concentration during eight days (see Figure 3(c)), which leads to both a recovery of the CTL number (that has been decreased by chemotherapy) and its following increase (see Figure 3(b)). Later on, the tumor steadily grows and the suppression of the immune functions takes place. Notice, that tumor growth rate at this stage is smaller than for $t < 8$ days. Thus, although this sequential regimen does not lead to the tumor regression it allows one to delay the tumor growth.

*4.4. Sequential immuno/chemotherapy*

Let us consider the following sequential regimen: one pulse of the IL-2 therapy, which is presented by setting in (3) $V_{I_2}(t) = 10$ MU/day for four days. For the next four days one pulse of chemotherapy is administered in dose $V_C(t) = 1$ in (5) per day for four days (Figure 3(d)). The dynamics of tumor cells is shown in Figure 3(a). As is seen, the result of this sequential regimen is worse in comparison with the previous case. Indeed, the IL-2 dosing leads to the increase of its concentration (Figure 3(c)) and, accordingly, to the increase of the CTL number (Figure 3(b)). However, the CTL have not enough time to achieve the magnitude sufficient to slow down the tumor evolution since their growth is abruptly stopped due to chemotherapy (see Figure 3(b)). Nevertheless, at the termination of course of treatment the CTL number again increases due to a sufficiently high concentration of IL-2. As a result, the tumor growth becomes slower reaching a dangerous size twenty days later than in the absence of therapy.

*4.5. The concurrent biochemotherapy*

The regimen of the sequential therapy is chosen to be the following: the chemotherapy in dose $V_C(t) = 1$ per day and the IL-2 therapy in dose $V_{I_2}(t) = 7$ MU/day are given simultaneously for four days. Since the concurrent chemoimmunotherapy is found to be less toxic in comparison with other sequential regimens (see, e.g., [15]), the dose of the IL-2 is selected to be approximately $3 \cdot 10^6$ units less than in subsections 4.3 and 4.4. As a result, the tumor cell dynamics becomes a little higher in comparison with the first sequential regimen in 4.3 during a period of time that is long enough, except for the initial interval of (0; 10) days (see Figure 3(a)). For this period of time, the tumor growth deceleration is more pronounced in comparison with the case of chemo/immuno sequence. Indeed, since chemotherapy and IL-2 therapy are used simultaneously, the tumor cells die under the action of both drug and the immune response recovered by IL-2 therapy. As is seen from Figure 3(b), the dynamics of CTL is similar to that without therapy. For the first six days the IL-2 concentration is higher than in the case of chemo/immunotherapy (Figure 3(c)). Thus, our simulations show that the stronger increase of the IL-2 concentration prevents the reduction in the CTL number caused by the chemical impact (unlike the first sequential regimen). In turn, for the next four days the IL-2 concentration becomes lower as compared with the case of chemo/immunotherapy. Therefore, one can conclude that the concurrent chemoimmunotherapy is more favorable in comparison with the regimen considered in the subsection 4.3.

*4.6. Vaccine therapy*

Cancer vaccines are considered as one of promising methods of immunotherapy. Using vaccine allows sensitizing the immune system to the presence of the certain forms of cancer. As a consequence, the immune system will be able to find and lyse tumor cells more effectively. When vaccine appears in the body the anti-tumor lymphocyte formation occurs. The efficacy of the vaccination depends on the following factors: (i) the number of tumor cells and their mitotic activity, (ii) the type of tumor, i.e. its histological structure, antigen structure, the number of HLA-A molecules expressed on the tumor cells and (iii) initial conditions of the immune system.

In this subsection, we consider a cancer vaccine consisting of four tumor-derived peptides with an adjuvant (see, for example, [33] for a current list of ongoing trials). As long as antigen/adjuvant complexes stimulate immune response to vaccine thereby enhancing immune reaction to patient's tumor cells, the effect of the vaccination can be taken into account through the model parameters. Therefore, in order to simulate vaccine therapy we change the values of four model parameters at the time of vaccination (in a similar manner as in [11]). The parameters that are sensitive to vaccination can be extracted from the experimental results obtained on mouse vaccine trials by Diefenbach et al [14]. Namely, we fitted the experimental curves produced by Diefenbach's data to our model and found the parameters that would change to reflect the administration of a therapeutic vaccine. They are $c_{\text{CTL}}$, the rate of inactivation of tumor cells by CTL; $e$, the rate of CTL proliferation induced by IL-2; $g$, the antigen presentation (the probability of interaction between helper T cell precursors and APC); and $j$, the rate of consumption IL-2 by CTL. Finally, to simulate vaccine therapy we alter the corresponding model parameters in the same direction as they change in Diefenbach's murine model [14] (cf. [11]). As a result, all four parameters ($c_{\text{CTL}}$, $e$, $g$, and $j$) are found to be increased.

We present here the results for vaccine therapy alone, so that we put $V_C(t)$, $V_{I_2}(t)$, $V_I(t)$ equal to zero as well as $C = I = 0$ in (1)-(5). The regimen of vaccination chosen for simulation of vaccine therapy is the following: the cancer vaccine administered once a week during 1—3, 5—7, 13, 27, 40, and 53 weeks, respectively [33]. We suppose that the vaccine is effective 83 days after the last injection. This value is not imperative. It seems plausible that this action may last even longer. We assume that at the expiration of this period the system parameters are restored to their initial values. As a result, tumor growth restarts. Therefore revaccination is required to avoid a disease recurrence.

From the above discussion it is clear that the values of parameters $c_{\text{CTL}}$, $e$, $g$, and $j$ will depend on the regimen of vaccination, i.e. on time. In other words, during the vaccine action we increase parameters by a certain percent value (see Table 3). Under these assumptions, the steady-state conditions for P1 become changed in such a way that the system (11)—(13) passes to the region II on the bifurcation diagram (see Fig.1 and Figure 4). Remember that in this region treatment outcome markedly depends on the initial tumor size and the immune system conditions. Figure 5 shows the results for two courses of the vaccination: the first one was administered without delay while the second one was administered 10 days later, when tumor cell population has reached a sufficiently large value to escape the immune response (Figure 5(a), 5b, 5c).

Let us first analyze the behavior of the system under the vaccine administered after surgery. For therapy without delay, the initial number of tumor cells is enough to induce the immune response. As is seen from Figure 5(c), the IL-2 concentration grows and, consequently, the CTL number is increased. The integral curves tend to the stable spiral point and the long tumor remission is observed (Figure 5(a)). Assumed 10-day delay is simulated by a time displacement $t \to t + 10$. In this case, the tumor has time to reach a sufficiently large size and both the IL-2 concentration and CTL number are decreasing (Figure 5(b) and 5c). The integral curves tend to the improper node, which means progressive tumor. The simulations show that the earlier the vaccination is administered the more effective it is for the cancer treatment.

Let us simulate the vaccine administered without previous surgery. Figure 6(a) shows that even for the therapy without delay the tumor regression does not occur and only some stunted tumor growth with lower saturation level is observed in comparison with the case without therapy. One can suppose that the saturation level without therapy corresponds to a dangerous tumor size in stage II of malignant process. Then the lower saturation level with vaccination may be considered as a steady state of a patient during the vaccine action (Figure 6(a)). In the case of 10-day delay, the tumor size almost reaches the therapeutic saturation level (Fig 6a). As is seen, the vaccine-mediated enhancement of the immune response prevents tumor growth to reach a dangerous size. Namely, after 15 days of growth the tumor curve goes slightly down and tends to the therapeutic saturation level. This does not mean, however, that a delay in the vaccination is not dangerous. In fact, as mentioned above, we do not take into account the angiogenesis, which begins at certain size of the tumor and provokes its further explosion [2,7]. In other words, the existence of the saturation level does not imply the termination of the tumor growth. Figs. 6b and 6c show dynamics of CTL and IL-2, respectively.

*4.7. Comparison with second and third groups of patients*

Table 4 summarizes the main findings of subsections 4.1—4.6 for the first group of patients as well as presents the results for two other groups. Let us compare three groups of patients. According to Table 4, for the second group of patients the IL-2 alone and IL-2 plus IFN-α therapies result in slower tumor expansion as compared to chemotherapy. After 6 weeks of the IL-2 alone therapy the tumor volume reaches almost the same value as in the case of chemotherapy. Results of chemo/immuno regimen are found to be similar to immuno/chemo and concurrent chemoimmunotherapy. This markedly differs from the first group where the increase of tumor size for chemo/immune sequence is smaller in comparison with the reverse sequence. Therefore, one can conclude that the dependence on the schedule is more pronounced in the first group. As a possible reason, the IL-2 therapy is less effective in the second group. This is a result of lower tumor antigen expression when stimulation of CTL proliferation by IL-2 becomes insufficient for effective recognition of tumor cells. Although the IL-2 plus IFN-α therapy looks more favorable for P2 in comparison with other therapies, one has to bear in mind that the vaccine therapy is less toxic. Let us consider the behavior of the second group of patients in response to the vaccination more detail. The increase of corresponding parameters for P2 during the vaccine action is shown in Table 3. As is seen, the suggested values differ from these for P1. Figure 4 shows that in this case the steady-state conditions do not change. This result looks rather unexpected. In fact, it means absence of the positive clinical response despite the fact that the immune reaction to the tumor is taken to be enhanced by the vaccine even better as compared with the first group. Indeed, we have intentionally taken the bigger relative growth of parameters *e*, *g*, and *j* for P2 in comparison with P1.

Figure 7 shows the results of the vaccination after surgery and Figure 8 shows the case without preliminary treatment. In the first case, using vaccine without delay allows stunting tumor growth and it reaches the therapeutic saturation level in 70 days. Besides, with vaccine the saturation level becomes lower than without therapy. It should be noted that the vaccine administered with 10-days delay is not effective because no deceleration of the tumor growth is observed (Figure 7(a)). The vaccination is ineffective when it is administered without preliminary surgery (Figure 8(a)).

It should be noted that the positive response to the treatment can also be described within our model. In order to show this possibility we consider the third group of patients (P3 in Table 2). In the absence of treatment the tumor grows to the dangerous size (see Figure 9). As is seen from Table 4, all of the considered therapeutic regimens result in tumor regression to the small volume that corresponds to the stable fixed point (spiral node) of the system. For sequential regimens the slower decreasing of the tumor volume is found as compared to IL-2 alone, IL-2 plus IFN-α and vaccine therapies. There is more pronounced regression of the tumor size in the cases of chemo/immune sequence and concurrent chemoimmunotherapy in comparison with the immune/chemo sequence. At the same time, we would like to mention more than 50% decrease of the tumor size that implies the effectiveness of all considered regimens. Thus, our simulations show that after cessation of therapy the tumor regresses.

The interesting results are obtained in the case of the vaccine therapy (see Figure 9). We consider the situation when under the vaccine therapy the system passes to the region III in Fig.1 (see Table 3). In this case, the effect of vaccine therapy does not depend on the initial tumor size and the immune system conditions. As a result, the time delay is out of importance (without angiogenesis taken into account). As is seen from Figure 9(a), the tumor cells population decreases to a small size. While after the termination of the vaccine action the tumor regrows, it nevertheless never exceeds the size $T_1$ (corresponding to the stable spiral for P3).

## 5. Conclusion

We have studied the effects of different treatment regimens on both the tumor growth and the immune response within the simple ODE model that describes tumor-immune dynamics with chemotherapy and immunotherapy. It is found that the regime of full regression of tumor is not admitted in our model. This conclusion is in agreement with some current clinical observations where recurrences of tumors are observed [26]. The bifurcation diagram for antigen presentation shows three main dynamical regimes. The region I reflects a progressive growth when the tumor is able to escape from the immune response. The region II describes two regimens of disease depending on both the initial tumor size and the condition of immune system: (i) the regression to small tumor when the dynamical equilibrium is established and (ii) a progressive tumor growth to the highest possible size. For the region III the decrease of the tumor size is found when the equilibrium between the tumor and the immune system is established.

In order to describe a possibility of different responses to treatment regimens, patients were conditionally divided in three generalized groups. Each group is characterized by specific tumor antigen expression, the strength of the immune response, and the reaction to vaccination. For patients with a weak immune response the vaccine therapy is found to be the most effective in comparison with other described treatments when used without time delay from a prescribed date of vaccination after surgery. This means that using vaccine gives the best results for patients with both small size of tumor and an immune system, which is not suppressed by tumor growth and able to respond to the vaccine. For the first group, the vaccine therapy is shown to be the only possible treatment allowing long tumor remission. Therefore, we note a promising effect of the vaccine treatment to improve immune response for this group. This qualitatively agrees with clinically observed results (see, e.g., [40,45,46,50]). For the second group all considered treatments result in progressive growth. However, the vaccine therapy without delay after surgery is expected to be more sparing. We observed that for patients with a strong immune response IL-2 alone, IL-2 plus IFN-$\alpha$ and the sequential chemoimmunotherapy could be used as a reasonable alternative to vaccination.

Our study shows that along with progressive disease the positive clinical responses to the treatment characterizing by a long remission of tumor growth are possible. This qualitatively agrees with modern clinical observations. Indeed, clinical trials of chemotherapy and sequential regimens for melanoma showed that along with progressive diseases the partial responses and even complete responses were possible within patient groups [26]. It was also shown in the trials of the vaccine therapy [45] that the CTL response to the vaccine by itself did not guarantee the tumor regression. For instance, for several patients the T cell response to the vaccine was found to be not strong enough to decrease the tumor size and, as a result, the tumor was progressing. At the same time, the tumor regression was observed for few patients with immune responses to the vaccine. In our model, these observations could be explained by both heterogeneity of the tumor antigen expression and patient-specific characteristics of immune response.

It should be stressed that all these predictions can be valid only for the description of early stages of the tumor growth when the processes of angiogenesis, invasion and metastasis are not of critical importance. As the next step, we plan to extend our simple ODE model to PDE model to include spatial components for distributions of cells and drugs. In this context, the presented analysis is of interest since it gives the underlying kinetics of more general PDE model. Notice also that toxicity of both chemotherapy and immunotherapy requires finding of an optimal treatment protocol. There are known some attempts to examine this problem within appropriate ODE models of tumor-immune dynamics (see e.g. [3,12,13]) and it would be interesting to consider it within our model. These studies are now in progress.

**Table captions**

**Table 1**. The parameter set for the first group of patients (P1).

| Parameter | Units | Description | Value | Source |
|---|---|---|---|---|
| $a$ | day$^{-1}$ | Tumor growth rate | 0.13 | Fit to data [21] |
| $b$ | cell$^{-1}$day$^{-1}$ | $a/b$ is a tumor carrying capacity | $3 \times 10^{-10}$ | Fit to data [21] |
| $c_{\text{CTL}}$ | cell$^{-1}$ day$^{-1}$ | Rate of tumor cells inactivation by CTL | $4.4 \times 10^{-9}$ | |
| $d$ | cell day$^{-1}$ | Rate of steady inflow of CTL | $7.3 \times 10^{6}$ | Estimated from [17] |
| $e$ | cell$^{-1}$ day$^{-1}$ | CTL proliferation rate induced by IL-2 | $9.9 \times 10^{-9}$ | |
| $f$ | day$^{-1}$ | CTL death rate | 0.33 | Estimated from [6] |
| $g$ | unit day$^{-1}$ | Antigen presentation | $1.6 \times 10^{7}$ | |
| $j$ | cell$^{-1}$ day$^{-1}$ | Rate of consumption of IL-2 by CTL | $3.3 \times 10^{-9}$ | |
| $k$ | cell$^{-1}$ day$^{-1}$ | Inactivation of IL-2 molecules by prostaglandines | $1.8 \times 10^{-8}$ | |
| $l$ | cell | Half-saturation constant | $3 \times 10^{6}$ | |
| $M_T^{\text{chemo}}$ | day$^{-1}$ | Tumor cell killing by chemotherapy | 0.9 | Taken from [11] |
| $M_L^{\text{chemo}}$ | day$^{-1}$ | CTL killing by chemotherapy | 0.6 | |
| $p$ | day$^{-1}$ | Decay rate of chemotherapy drug | 6.4 | Estimated from [29] |
| $q$ | day$^{-1}$ | Decay rate of therapeutic IFN−α | 1.7 | Estimated from [43] |
| $T_0 = 10^8$ cells | $L_0 = 9 \times 10^7$ cells | $I_{20} = 2 \times 10^7$ units | $I_0 = 10^7$ units | |

**Table 2.** The parameter sets for the second (P2) and the third (P3) groups of patients.

| Parameter | Value | |
|---|---|---|
| | P2 | P3 |
| $a$ | 0.13 | 0.13 |
| $b$ | $3 \times 10^{-10}$ | $3 \times 10^{-10}$ |
| $c_{\text{CTL}}$ | $3.3 \times 10^{-9}$ | $5.5 \times 10^{-9}$ |
| $d$ | $7.3 \times 10^{6}$ | $7.3 \times 10^{6}$ |
| $e$ | $9.6 \times 10^{-9}$ | $1.0 \times 10^{-8}$ |
| $f$ | 0.33 | 0.33 |
| $g$ | $1.4 \times 10^{7}$ | $2.4 \times 10^{7}$ |
| $j$ | $2.9 \times 10^{-9}$ | $3.7 \times 10^{-9}$ |
| $k$ | $1.5 \times 10^{-8}$ | $2.1 \times 10^{-8}$ |
| $l$ | $3 \times 10^{6}$ | $3 \times 10^{6}$ |
| $M_T^{\text{chemo}}$ | 0.9 | 0.9 |
| $M_L^{\text{chemo}}$ | 0.6 | 0.6 |
| $p$ | 6.4 | 6.4 |
| $q$ | 1.7 | 1.7 |

**Table 3.** The percent increase of the parameters sensitive to the vaccine

| Parameter | Increase (in %) | |
|---|---|---|
| | P1 | P2, P3 |
| $c_{CTL}$ | 20 | 10 |
| $e$ | 15 | 20 |
| $g$ | 20 | 30 |
| $j$ | 30 | 50 |

**Table 4.** Effects of different treatments for three groups of patients[*].

| | **P1** | | **P2** | | | | **P3** | | | |
|---|---|---|---|---|---|---|---|---|---|---|
| | $t_0$ | $\Delta T$ | $t_0$ | $\Delta T$ | $t_0$ | $\Delta T$ | $t_0$ | $\Delta T$ | $t_0$ | $\Delta T$ |
| **Chemotherapy** | 6 | ↑ 13.4 | 3 | ↑ 11.4 | 6 | ↑ 20 | 3 | ↑ 0.1 | 6 | ↓ 0.43 |
| **IL-2 alone** | | ↓ 0.34 | | ↑ 7.5 | | ↑ 21 | | ↓ 0.82 | | ↓ 0.92 |
| **IL-2 plus IFN-α** | | ↓ 0.42 | | ↑ 3.9 | | ↑ 10 | | ↓ 0.95 | | ↓ 0.93 |
| **Chemo/immune** | | ↑ 9.92 | | ↑ 9.3 | | ↑ 23.9 | | ↓ 0.68 | | ↓ 0.88 |
| **Immune/chemo** | | ↑ 12.8 | | ↑ 10.2 | | ↑ 24.3 | | ↓ 0.48 | | ↓ 0.88 |
| **Chemo + immune** | | ↑ 10.9 | | ↑ 9.5 | | ↑ 24 | | ↓ 0.63 | | ↓ 0.88 |
| **Vaccine therapy without 10 day delay** | | ↓ 0.25 | | ↑ 8.92 | | ↑ 18.7 | | ↓ 0.91 | | ↓ 0.91 |

[*]The therapies for P1 and P2 are administered after surgery, and for P3 without previous treatment. The arrow ↑ means the increase of the tumor size, ↓ — the decrease of the tumor size. The change of tumor size is presented as $\Delta T = (T(t_0) - T(0))/T(0)$. $t_0$ — the time after the start of treatment, weeks For P1 and P2 the initial conditions are $T(0) = 8 \times 10^6$ cells, $L(0) = 2.25 \times 10^7$ cells, $I(0) = 2.4 \times 10^7$ cells, for P3 — $T(0) = 3 \times 10^7$ cells, $L(0) = 3.45 \times 10^7$ cells, $I(0) = 1.7 \times 10^7$ cells

## Figure captions

**Figure 1.** A bifurcation diagram varying the antigen presentation ($h_6$). For $h_6 < h_{6min}$ there is only one steady state — improper node (region I). When $h_{6min} < h_6 < h_{6max}$ there are two stable steady states — improper node and spiral node as well as an unstable (saddle) point (region II). For $h_6 > h_{6max}$ only one steady state, the spiral node remains (region III).

**Figure 2.** Human data, group P1. Effects of chemo-, IL-2 and IL-2 plus IFN−α therapies on tumor and immune response dynamics. (a) tumor cells, (b) cytotoxic T cells, and (c) IL-2 vs. time. (d) shows drug administration pattern: nine doses, strength $V_C(t) = 1$, 1 day per dose on a 5 day cycle, and IFN−α administration pattern: four doses, strength $V_I(t) = 5$ MU/day, 4 days per dose on a 10 day cycle. IL-2 is administered with four doses of strength $V_{I_2}(t) = 10$ MU/day, 4 days per dose on a 10 day cycle. Initial conditions: $8\times10^6$ tumor cells, $2.25\times10^7$ cytotoxic T lymphocytes, $2.4\times10^7$ IL-2 units.

**Figure 3.** Human data, group P1. Effects of one pulse of chemotherapy followed immediately by one pulse of IL-2-therapy, one pulse of IL-2 therapy followed immediately by one pulse of chemotherapy, and concurrent chemoimmunotherapy. (a) tumor cells, (b) cytotoxic T cells, and (c) IL-2 vs. time. (d) shows drug administration pattern: one pulsed dose of chemotherapy, strength $V_C(t) = 1$ per day, 4 days per dose for sequential chemo/immunotherapy (dotted line), sequential immuno/chemotherapy (dash-dot line), and concurrent chemoimmunotherapy (gray line). IL-2 administration pattern: one pulsed dose of strength $V_{I_2}(t) = 10$ MU/day, 4 days per dose after chemotherapy (chemo/immunotherapy sequence) or before chemotherapy (immuno/chemotherapy sequence) and $V_{I_2}(t) = 7$ MU/day for four days simultaneously with chemotherapy (concurrent biochemotherapy). Initial conditions: $8\times10^6$ tumor cells, $2.25\times10^7$ cytotoxic T lymphocytes, $2.4\times10^7$ IL-2 units.

**Figure 4.** Bifurcation diagrams showing the effect of vaccine therapy on anti-tumor immune response dynamics for P1 and P2.

**Figure 5.** Human data, group P1. Effects of vaccine administered after surgery without delay and with delay for 10 days. (a) tumor cells, (b) cytotoxic T cells, and (c) IL-2 vs. time. Initial conditions: $8\times10^6$ tumor cells, $2.25\times10^7$ cytotoxic T lymphocytes, $2.4\times10^7$ IL-2 units.

**Figure 6.** Human data, group P1. Effects of vaccine administered without previous treatment and with delay for next 10 days. (a) tumor cells, (b) cytotoxic T cells, and (c) IL-2 vs. time. Initial conditions: $3\times10^7$ tumor cells, $1.35\times10^7$ cytotoxic T lymphocytes, $1.8\times10^7$ IL-2 units.

**Figure 7.** Human data, group P2. Effects of vaccine administered after surgery without delay and with delay for 10 days. (a) tumor cells, (b) cytotoxic T cells, and (c) IL-2 vs. time. Initial conditions: $8\times10^6$ tumor cells, $2.25\times10^7$ cytotoxic T lymphocytes, $2.4\times10^7$ IL-2 units.

**Figure 8.** Human data, group P2. Effects of vaccine administered without previous treatment and with delay for next 10 days. (a) tumor cells, (b) cytotoxic T cells, and (c) IL-2 vs. time. Initial conditions: $3\times10^7$ tumor cells, $1.35\times10^7$ cytotoxic T lymphocytes, $1.8\times10^7$ IL-2 units.

**Figure 9.** Human data, group P3. Effects of vaccine administered without previous treatment. (a) tumor cells, (b) cytotoxic T cells, and (c) IL-2 vs. time. Initial conditions: $3\times10^7$ tumor cells, $3.45\times10^7$ cytotoxic T lymphocytes, $1.7\times10^7$ IL-2 units.

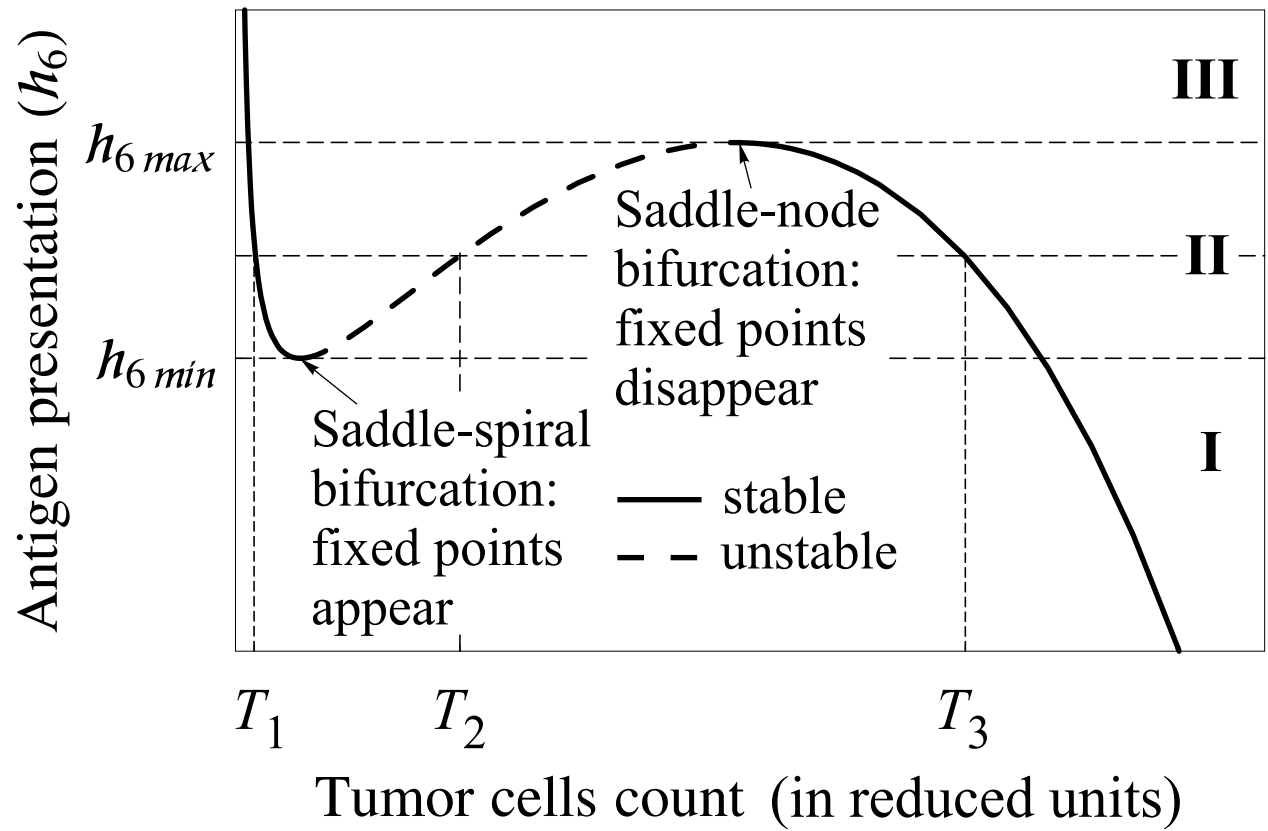

**Figure 1**

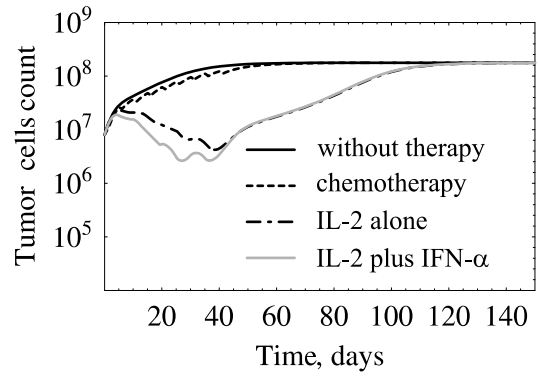
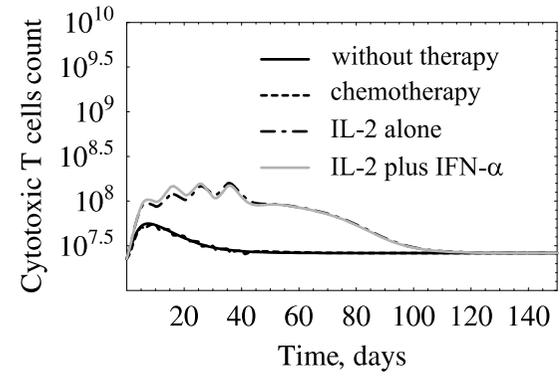
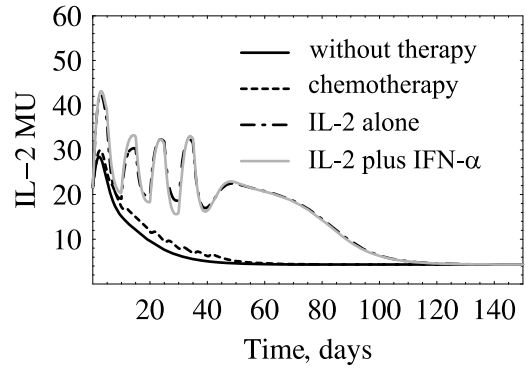
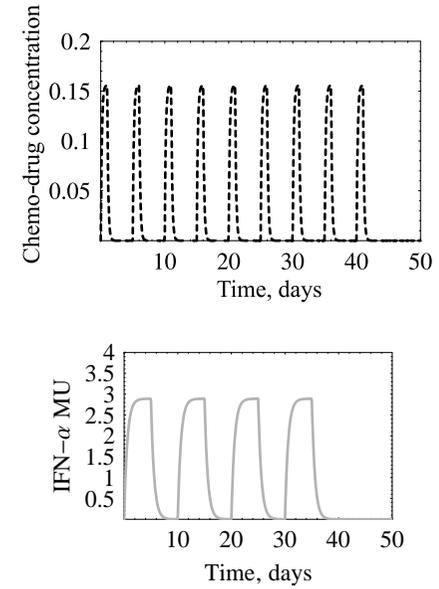

**Figure 2**

a 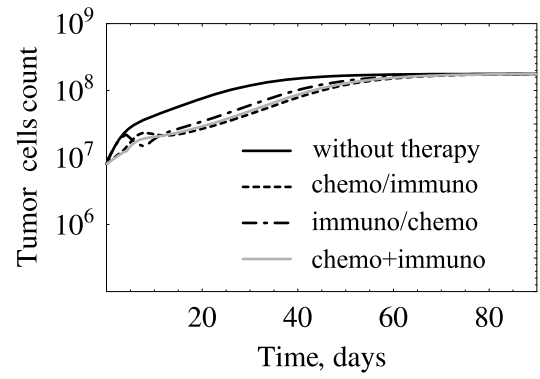 b 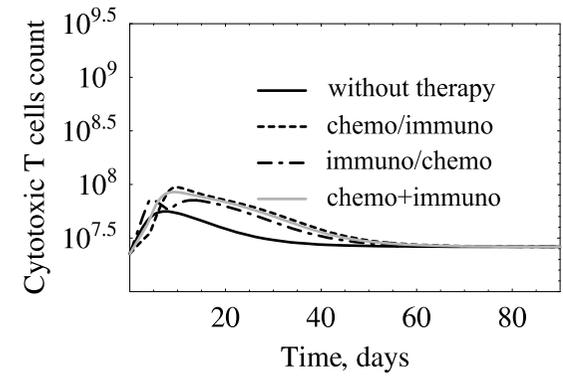

c 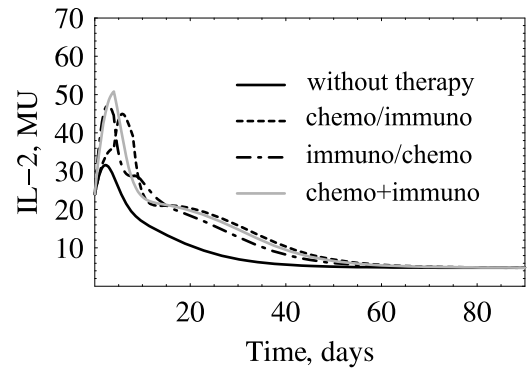 d 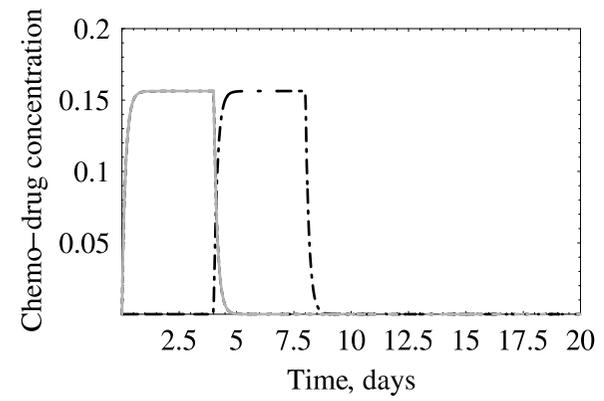

**Figure 3**

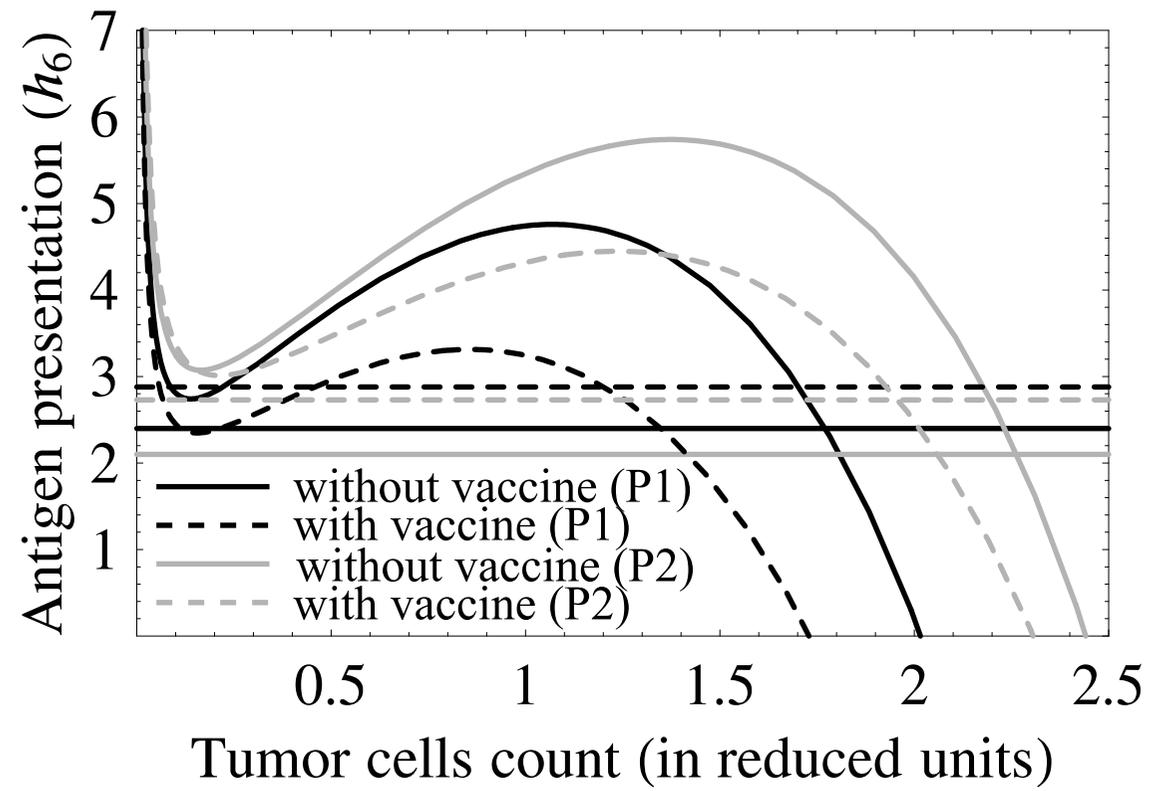

**Figure 4**

a 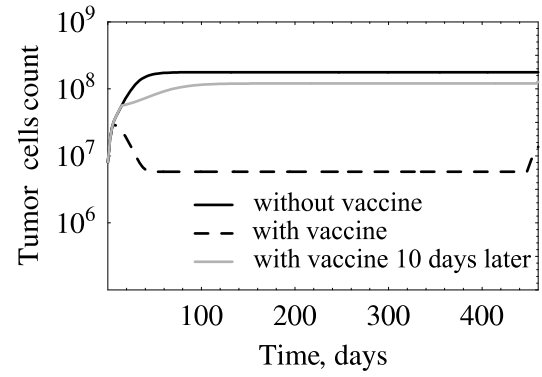

b 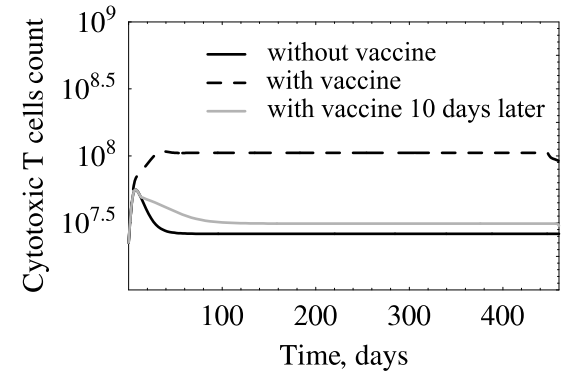

c 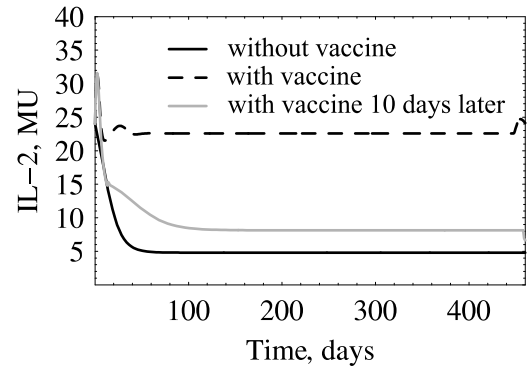

**Figure 5**

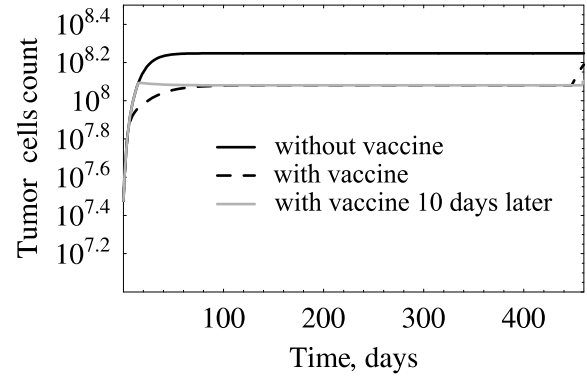
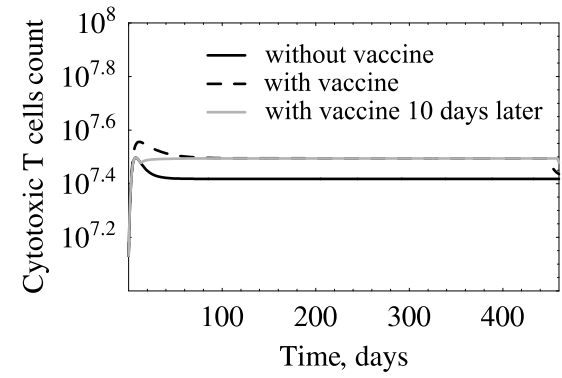
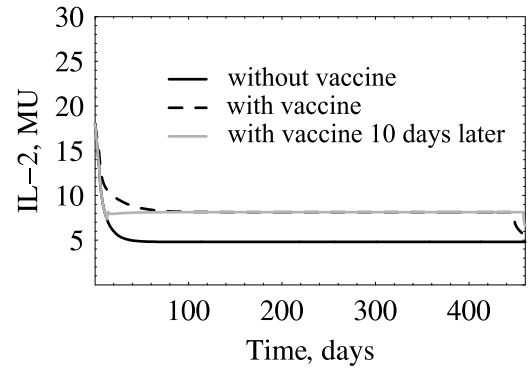

**Figure 6**

a 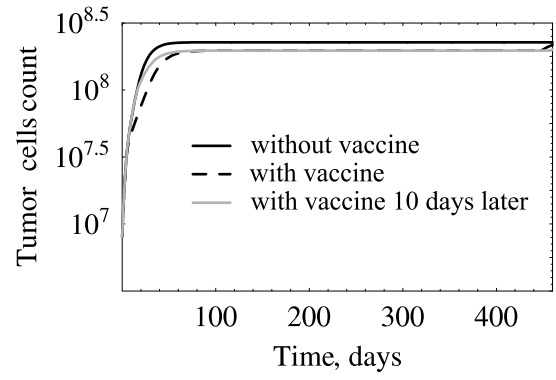

b 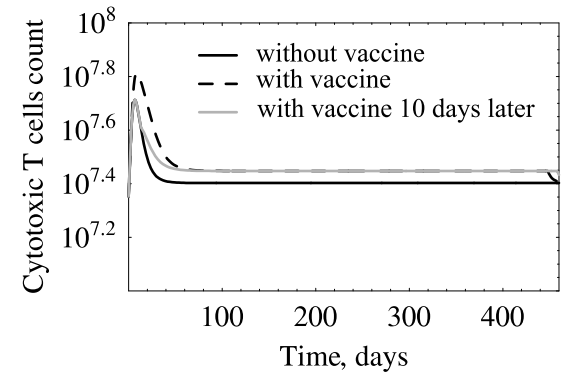

c 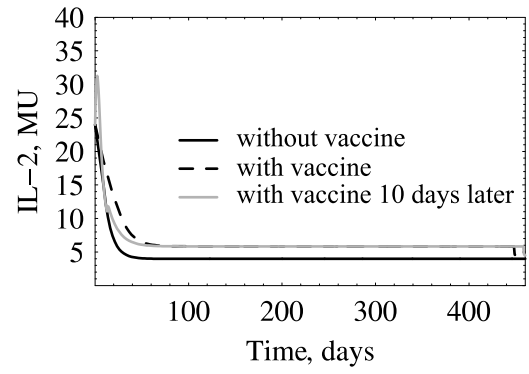

**Figure 7**

a 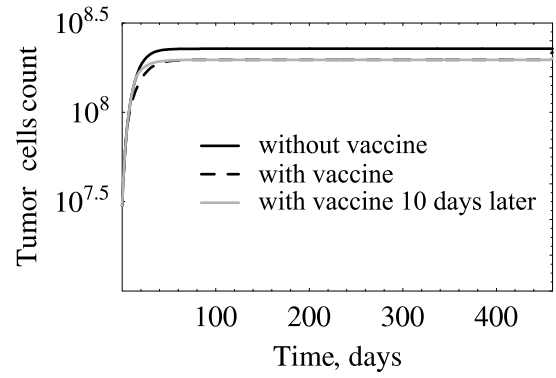

b 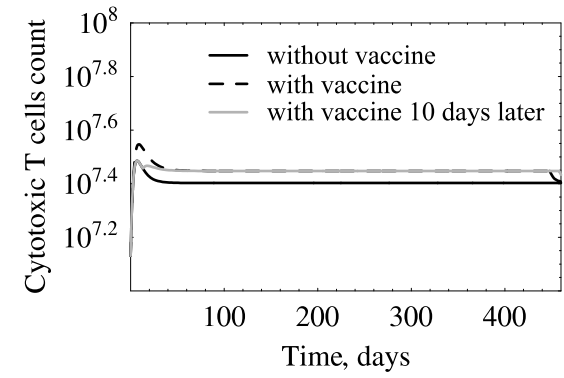

c 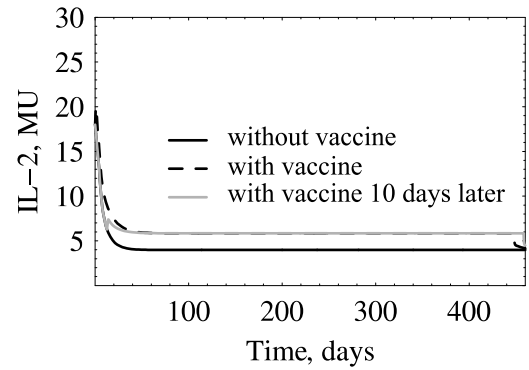

**Figure 8**

a
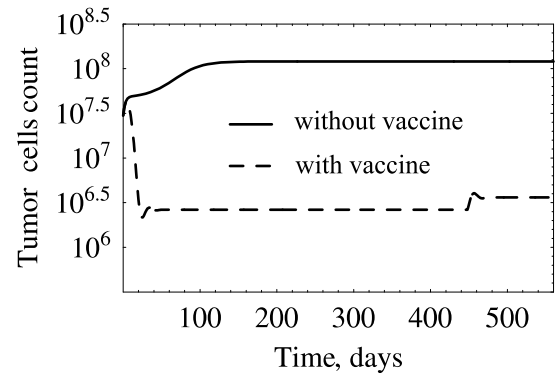

b
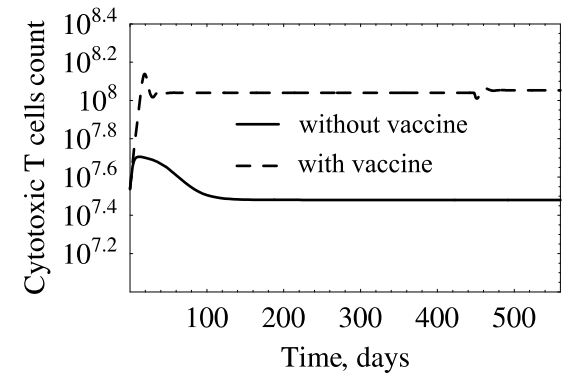

c
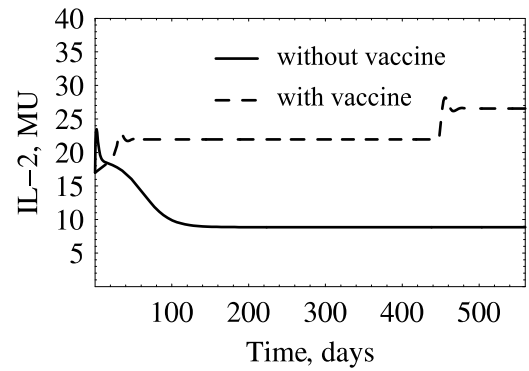

**Figure 9**